\documentclass[aps,prd,onecolumn,nofootinbib,amsmath,amssymb,floatfix,letterpaper,notitlepage]{revtex4-1}
\usepackage{graphicx}
\usepackage[hidelinks]{hyperref}
\usepackage{color}
\usepackage{amsmath,amssymb}
\usepackage{bm}
\usepackage[utf8]{inputenc}
\usepackage{subcaption}
\usepackage{tikz}
\usetikzlibrary{decorations.pathmorphing}
\usetikzlibrary{arrows}

\newcommand{\comment}[1]{}
\renewcommand{\d}{\partial}
\newcommand{\beq}{\begin{equation}}
\newcommand{\eeq}{\end{equation}}
\newcommand{\bea}{\begin{eqnarray}}
\newcommand{\eea}{\end{eqnarray}}
\newcommand{\bsp}{\begin{split}}
\newcommand{\esp}{\end{split}}

\newcommand{\vx}{\vec x}
\newcommand{\vv}{\vec v}

\newcommand{\vk} {{\boldsymbol k}}
\newcommand{\vp} {{\boldsymbol p}}
\newcommand{\vps}{\vec \psi}

\newcommand{\vq}{\vec q}
\newcommand{\confh}{\mathcal{H}}

\renewcommand{\vec}[1]{\bm{#1}}

\definecolor{darkgreen}{RGB}{0,120,0}

\begin{document}
\title{Lagrangian Formulation of the Eulerian-EFT }
\author{Matias Zaldarriaga}
\email{matiasz@ias.edu}
\affiliation{Institute for Advanced Study, Princeton, NJ, USA}
\author{Mehrdad Mirbabayi}
\affiliation{Institute for Advanced Study, Princeton, NJ, USA}
\begin{abstract}
We study the counter terms in the Eulerian version of the EFT of Large Scale Structure. We reformulate the equations to solve for the displacement of fluid elements as a bookkeeping variable and study the structure of the counter terms in this formulation. We show that in many cases the time dependence of the amplitude of the counter terms is irrelevant, as solutions obtained for various time dependences differ by terms that can be reabsorbed by higher order counter terms. We show that including all effects due to the non-locality in time and the time dependence of the counter terms there are six  new parameters relevant for the two loop power spectrum calculation. We give explicit expressions for all these terms and study the contributions to them from large and small modes. We show that the shape of all these terms is very similar. 
\end{abstract}
\maketitle
\section{Introduction}

Understanding the development of structure is one of our main tools to constrain the history of the Universe, its matter content and the laws that govern its evolution. On large scales perturbations are small and are thus amenable to analytical techniques. 
Standard Perturbation Theory (SPT, see e.g.~\cite{Bernardeau:2001qr}) has allowed cosmologists to gain important insights about gravitational clustering.  However it was soon realized that in order to extend the validity of the theory beyond the leading order, one needs do go beyond SPT. Through the years, different proposals have been made, e.g.~\cite{Crocce:2005xy,Bernardeau2008,Taruya2012,Blas:2013aba}. In recent years an Effective Theory for Large Scale Structure (EFT of LSS) \cite{Baumann:2010tm,Carrasco:2012cv} 
has been introduced as a way to systematically incorporate the effects of the small scale non-linear dynamics on the larger perturbative scales. 

Since the original papers \cite{Baumann:2010tm,Carrasco:2012cv}, many aspects of the EFT of LSS have been explored. One-loop calculations were presented in \cite{Hertzberg2014,Pajer:2013jj,Mercolli:2013bsa,Carroll2014,Senatore:2014via,Senatore2014a,Foreman:2015uva} for the power spectrum while \cite{Baldauf:2015qfa, Angulo:2014tfa} considered the matter bispectrum. Two loop computations were presented in \cite{Carrasco:2013sva,Carrasco:2013mua,Foreman:2015lca,Baldauf:2015aha}  and \cite{Assassi2015} looked at non-Gaussian effects. Systematics of renormalization was studied in \cite{Abolhasani}. The Lagrangian space formulation of the EFT of LSS was studied in \cite{Porto:2013qua,Vlah2015}. Bias and baryonic effects were considered in \cite{McDonald2006,McDonald2009,Schmidt2013,Assassi2014a,Senatore2014,Mirbabayi2014,Lewandowski2015,Angulo2015}. Finally detailed comparison with numerical simulations at the field level were presented in \cite{Baldauf:2015zga,Baldauf:2015tla}.

In the EFT approach, the effect of small scales on large ones is encoded in an effective stress tensor. Although the different operators or counter terms that contribute to this effective tensor can be listed once the desired precision is prescribed, the normalization of these terms is free. The free parameters need to be calibrated either using observations or simulations. The number of terms grow quickly as the desired precision increases. 

For one loop calculations of the matter power spectrum only one parameter is relevant for cosmologies similar to ours. Three additional parameters are in principle necessary to predict the one loop bispectum. Even more parameters are needed, at least in principle for the two-loop power spectrum. In addition this normalization coefficients for the different operators should be taken as time dependent. Another complication is that the EFT is a  non-local in time theory so one should really write the operators contributing to the stress tensor as integrals along the fluid trajectory, or equivalently allow any number of convective derivatives acting on the building blocks used to construct the stress tensor without any suppression. It seems that the number of free parameters grows too quickly as one strives for more precision. It is important to point out however that the corrections induced by these different terms are usually quite similar so in practice one can restrict the number substantially \cite{Foreman:2015lca,Baldauf:2015aha}. The fact that many of them are not identical traces to the fact that the primordial power spectrum of fluctuations is not a power-law. 

In this paper we want to study the structure of these counter terms in some detail.  To do so we will reformulate the Eulerian equations so as to solve for the trajectory of fluid elements which will satisfy the standard equations of Lagrangian perturbation theory except for the new terms coming from the stress tensor source. We found this new formulation convenient to easily include the effects associated with the non-locality in time. We also found that the vertices induced by the new counter terms are slightly simpler than in the standard formulation. 

It is easy to show that for an $n^{th}$ order counter term such that the divergence of the induced stress tensor is the gradient of a scalar $\partial_j \tau^{ij} = \d_i \chi$, the results up to the $n+2$ order are independent of the time dependent normalization of the term. Namely, the results obtained with various time dependences differ in ways that can be reabsorbed by changing higher order counter terms. At the order required for a two loop power spectrum calculation there is only one counter term for which $\partial_j \tau^{ij}$ has a curl. 

We explicitly calculated all the contributions for counter terms to the two loop power spectrum and found:
\bea\label{summary}
\langle \delta^{ct} \delta^{SPT} \rangle &=&  l_1^2 k^2 \left(P(k)+P_{22}(k)+2 P_{13}(k)+{\bar P}_{22}(k)+2 {\bar P}_{13}(k)\right) \nonumber \\
&+& 2 k^3 \int {dr r^2 \over (2 \pi)^2 } P(k r) \int_{-1}^1 dx P(k \sqrt{y(r,x)}) \left[{3 \over 14} {(1-x^2)\over y(r,x)} +  {1 \over 2} {(1-r x) x\over r y(r,x)}\right] \nonumber \\
&\times& \left[ k^2 l_{2a}^2  +  k^2  { l}_{2b}^2   {(1-x^2)\over y(r,x)} + k^2  {l}_{2c}^2 {(1-r x) (x-r) x\over r y(r,x)}  \right ] \nonumber \\
&+&   P(k) 
 2 k^3 \int {dr r^2 \over (2 \pi)^2 } P(k r)   \nonumber \\
 &\times&  \int_{-1}^1 dx  \left[k^2  \bar {l}_{2c}^2 {x^2 (1-r x) (1-x^2) \over y(r,x)} + k^2 l_{3a}^2  {(1-x^2)^2 \over y(r,x)} +  k^2  { l}_{3b}^2   {r x (1-x^2)^2 \over y(r,x)}  \right ]. 
\eea
Here $P(k)$ is the linear power spectrum, $y(r,x)=1+r^2-2rx$ and
\bea
{P}_{22}(k)&=& \int_p  2 F_2^2(\vp,\vk-\vp) P(p) P(|\vk-\vp|)\nonumber \\
{P}_{13}(k)&=& P(k) \int_p 3 F_3(\vk,\vp,-\vp)  P(p) \nonumber \\
{\bar P}_{22}(k)&=& \int_p 4 F_2^2(\vp,\vk-\vp) {\vp\cdot(\vp-\vk)\over k^2} P(p) P(|\vk-\vp|) \nonumber \\
{\bar P}_{13}(k)&=& P(k) \int_p 3 F_3(\vk,\vp,-\vp) {p^2 \over k^2}  P(p),
\eea
where $F_2$ and $F_3$ are the standard SPT kernels. There are six free parameters in this expression in addition to the lowest order one present for the one loop calculation. One should also consider adding a higher derivative counter term $k^4 l_4^4 P(k)$. Finally there are many additional terms one could add that contribute as $k^2 l^2 P(k)$ but this $k$ dependence was already fixed by the one loop term (proportional to $l_1^2$) thus they need not be included as long as the piece of the two-loop SPT result that scales this way is also discarded. Otherwise they must be picked to identically cancel this contribution and thus do not constitute a free parameter. 
Equation (\ref{summary}) includes all effects associated with the non-locality in time and the time dependence of the normalization coefficients. Of all these counter terms, two of the parameters $l_{2c}$ and $\bar l_{2c}$ originate from the same counter term, the quadratic stress tensor counterterm whose divergence is not a pure gradient. Their ratio would be known if the time dependent normalization of the quadratic stress tensor counterterm was known. The shape of these terms is shown to be very similar.

\section{The EFT of LSS: Lagrangian formulation}

In the EFT of LSS one sets to solve perturbatively the following equations:
\bea
\partial_\tau \delta + \partial_i[(1+\delta)v^i]&=&0  \;, \nonumber  \\
\partial_\tau v^i + \confh v^i + \partial^i \phi +  v^j \partial_j v^i &=& - {1 \over  \rho} \partial_j \tau^{i j} \equiv - {1 \over  (1+\delta)} \partial_j \tilde\tau^{i j}  \;, \label{eqspt}\\
\Delta \phi &=& {3 \over 2} \confh^2 \Omega_\text{m} \delta \;.\nonumber
\eea
These equations differ from those of standard perturbation theory (SPT) due to the addition of a new source,  a stress tensor source $\tau_{ij}$ in the Euler equation. This source arises due to the effect of small scales where the perturbative solution of SPT is not applicable.  What the EFT of LSS provides is an organizing framework for how to model this source, namely a list of terms with their associated free parameters that need to be introduced to achieve a desired accuracy. 

EFT is a theory for the long wavelength modes that is not local in time, as both long and short modes evolve with the same timescale, the Hubble scale. This long memory of the short modes implies that the operators introduced  to model  $\tau_{ij}$  need to be expressed as integrals along the fluid trajectory \cite{Carrasco:2013mua,Abolhasani}. Thus we start by defining the fluid trajectory.

Consider a fluid element that at time $\tau_0$ is located at position $\vx_0$. Its position at time $\tau$ is given by $\vx_{fl}(\vx_0,\tau_0;\tau)$ which satisfies:
\beq
\vx_{fl}(\vx,\tau;\tau')+\int_{\tau'}^{\tau} d\tau'' \vv(\vx_{fl}(\vx,\tau;\tau''),\tau'') = \vx.
\eeq 
It is convenient to define the (Lagrangian) location of a fluid element at the initial time $\vq(\vx,\tau)\equiv\vx_{fl}(\vx,\tau;0)$. The displacement field is then defined as
\beq
\vps(\vx,\tau) = \vx- \vq(\vx,\tau).
\eeq
To make contact with the standard notation of the Lagrangian perturbation theory it is useful to invert the function $\vq(\vx,\tau)$ and express the displacement field in Lagrangian coordinates 
\beq
\vps(\vq,\tau) = \vx(\vq,\tau)- \vq = \int_0^\tau d\tau' \vv(\vx(\vq,\tau'),\tau').
\eeq
From this definition it follows that
\beq
\dot \vps(\vq,\tau) \equiv{\partial \over \partial \tau}\vps(\vq,\tau) = \vv(\vx(\vq,\tau),\tau),
\eeq
and
\beq
\ddot \vps=D_\tau\vv \equiv ({\partial \over \partial \tau} + \vv\cdot\nabla)\vv,
\eeq
where we introduced the convective derivative $D_\tau$. Thus the Euler equation becomes:
\beq\label{xfleq}
\ddot \psi^{i} + \confh \dot \psi^{i} + \frac{\partial}{\d x^i} \phi =
- {1 \over (1+\delta) } \frac{\partial}{\d x^j} \tilde\tau^{i j},
\eeq
One can decompose $\psi^{i}$ into a gradient and a curl piece. We will now obtain equations for each of these pieces.  The treatment follows closely the literature of Lagrangian perturbation theory, we are just adding the new terms coming from the stress-tensor counter terms. 

\paragraph*{Scalar component of the displacement:}
For the scalar equation we take the divergence of the acceleration equation, yielding:
\beq\label{xfleulerian}
\frac{\partial}{\partial x_i}\left(\ddot \psi^{i}+\confh \dot \psi^{i}\right)=-\frac{3}{2}\confh^2 \Omega_\text{m}\delta - \frac{\partial}{\partial x_i} \Big({1 \over (1+\delta) } \frac{\partial}{\d x^j} \tilde\tau^{i j}\Big).
\eeq
From the mapping between Lagrangian and Eulerian space $\vec x=\vec q+\vec \psi$ and mass conservation
\beq\label{measure}
\bar \rho \ d^3\vq = \bar\rho (1+\delta) \ d^3\vx,
\eeq
we have that $1+\delta=1/J$, where $J$ is the determinant of the Jacobian matrix
\beq
A_{ij}\equiv\frac{\partial x_i}{\partial q^j}=\delta_{ij}+\psi_{i,j}
\eeq
where $\delta_{ij}$ is the Kronecker delta (not to be confused with density contrast). Spatial indices are always raised and lowered using $\delta_{ij}$ and its inverse. When convenient we will use a comma to denote spatial derivatives. Furthermore if not explicitly mentioned partial derivatives are with respect to the Lagrangian coordinate $q$. 
The determinant is:
\begin{align}
J=\text{Det}[A]=&\frac{1}{3!}\epsilon_{ijl}\epsilon_{stu}A_{is}A_{jt}A_{lu}\nonumber\\
=&1+\psi_{i,i}+\frac12 \left(\psi_{i,i}\psi_{j,j}-\psi_{i,j}\psi_{j,i}\right)+\frac{1}{3!}\epsilon_{ijl}\epsilon_{stu}\psi_{i,s}\psi_{j,t}\psi_{l,u}\label{eq:determinant}\\
\equiv&1+\mathcal{K}+\mathcal{L}+\mathcal{M}.\nonumber
\end{align}
We will need the mapping from Eulerian to Lagrangian derivatives
\beq
\frac{\partial}{\partial x_j}=A_{ij}^{-1}\frac{\partial}{\partial q_i}
\eeq
with
\begin{align}
A_{ij}^{-1}=\frac{C_{ij}}{\text{Det}[A]}=&\frac{1}{2!\; \text{Det}[A]}\epsilon_{jmn}\epsilon_{ist}A_{ms}A_{nt}\\
=&\frac{1}{J}\biggl(\delta_{ij}+(\delta_{ij}\delta_{nt}-\delta_{jt}\delta_{in})\psi_{n,t}+\frac{1}{2}\epsilon_{jmn}\epsilon_{ist}\psi_{m,s}\psi_{n,t}\biggr).
\end{align}
For the equation of motion we now have
\beq
J A_{ki}^{-1}\left(\ddot \psi_{i,k}+\confh \dot \psi_{i,k}\right)=\frac{3}{2}\Omega_\text{m}\mathcal{H}^2\left(\mathcal{K}+\mathcal{L}+\mathcal{M}\right) - J A_{ki}^{-1}[J A_{lj}^{-1}\tilde\tau_{i j,l} ]_{,k}.
\eeq
The combination
\begin{align}
J A_{ij}^{-1}={C}_{ij}=&\frac{1}{2!}\epsilon_{jmn}\epsilon_{ist}A_{ms}A_{nt}\\
=&\biggl(\delta_{ij}+(\delta_{ij}\delta_{nt}-\delta_{jt}\delta_{in})\psi_{n,t}+\frac{1}{2}\epsilon_{jmn}\epsilon_{ist}\psi_{m,s}\psi_{n,t}\biggr)
\end{align}
only contains linear and quadratic terms. Furthermore, it is easy to show that
\beq
{ C_{lj,l}}=0
\eeq
which implies that:
\beq
C_{lj}{\cal O}_{,l} = (C_{lj}{\cal O})_{,l}
\eeq
for any function ${\cal O}(\vq)$. 
Thus we can rewrite the equation of motion as:
\beq\label{eqlag}
C_{ki}\left(\psi_{i,k}^{\prime\prime}+(1+{\dot \confh \over \confh})  \psi_{i,k}^{\prime}\right)=\frac{3}{2}\Omega_\text{m}\left(\mathcal{K}+\mathcal{L}+\mathcal{M}\right) - C_{ki} \bar\tau_{i j,kj},
\eeq
where we have introduced $^\prime = d/d\ln a$ and defined $\confh^2 \bar\tau_{i j} = C_{jl} \tau_{i l}$. 
In addition to the vertices induced by the stress tensor counter term, this equation has quadratic and cubic vertices. In the Eulerian version of the equation (\ref{xfleulerian}), for each counter term in $\tau^{ij}$ the $(1+\delta)^{-1}$ generated an infinite set of associated counter terms at higher order. This complication is gone in this formulation. 

For simplicity we will consider the equations for $\Omega_m=1$. In that case they become:
\beq\label{diveq}
C_{ki}\left(\psi_{i,k}^{\prime\prime}+{1\over 2}  \psi_{i,k}^{\prime}\right)=\frac{3}{2}\left(\mathcal{K}+\mathcal{L}+\mathcal{M}\right) - C_{ki}\bar\tau_{i j,kj}. 
\eeq
In the absence of the stress tensor source, this equations can be solved perturbatively following the standard practice by proposing a solution of the form:
\beq
\psi_{i} =\sum_n a^n \psi_{i}^{(n)}.
\eeq
The equation of motion leads to the following recursive solution:
\beq\label{diveqpert}
\begin{split}
\left(n^2+\frac{n}{2}\right) \psi_{i,i}^{(n)}=&\frac{3}{2}\left(\mathcal{K}^{(n)}+\mathcal{L}^{(n)}+\mathcal{M}^{(n)}\right)-2\sum_{m=1}^{n-1}\left(m^2+\frac{m}{2}\right)\mathcal{L}^{n-m,m}\\
&-3\sum_{m=1}^{n-2}\sum_{k+l=n-m\atop{k>0,l>0}}\left(m^2+\frac{m}{2}\right)\mathcal{M}^{k,l,m}\; .
\end{split}
\eeq
Where we introduced:
\bea
\mathcal{L}^{n,m}&=&\frac{1}{2!} \left(\psi_{i,i}^{(n)}\psi_{j,j}^{(m)}-\psi_{i,j}^{(n)}\psi_{j,i}^{(m)}\right) \nonumber \\
\mathcal{M}^{n,m,k}&=&\frac{1}{3!}\epsilon_{ijl}\epsilon_{stu}\psi_{i,s}^{(n)}\psi_{j,t}^{(m)}\psi_{l,u}^{(k)}.
\eea
If we introduce $g(n)= n^2+\frac{n}{2}$ we can write:
\beq
\begin{split}
\psi_{i,i}^{(n)}=&
-\sum_{n_1+n_2=n\atop{n_1,n_2>0}}\left(\frac{2 g(n_1) - 3/2}{g(n)-3/2}\right)\mathcal{L}^{n_1,n_2}\\&-\sum_{n_1+n_2+n_3=n\atop{n_1,n_2,n_3>0}}\left(\frac{3 g(n_1)-3/2}{g(n)-3/2}\right)\mathcal{M}^{n_1,n_2,n_3}\\
=&-\sum_{n_1+n_2=n\atop{n_1,n_2>0}}\left(\frac{2 g(n_1)- 3/2}{g(n)-3/2}\right)\frac{1}{2!}\epsilon_{mis}\epsilon_{mjt}\psi_{i,j}^{(n_1)}\psi_{s,t}^{(n_2)}\\&-\sum_{n_1+n_2+n_3=n\atop{n_1,n_2,n_3>0}}\left(\frac{3 g(n_1)- 3/2}{g(n)-3/2}\right)\frac{1}{3!}\epsilon_{i_1i_2i_3}\epsilon_{j_1j_2j_3}\psi_{i_1,j_1}^{(n_1)} \psi_{i_2,j_2}^{(n_2)} \psi_{i_3,j_3}^{(n_3)}
\; .
\label{eq:rec_count_scalar}
\end{split}
\eeq

We now consider the the case with the stress tensor source.  We will assume that $\bar\tau^{i j}\propto a^\gamma$ with an unknown $\gamma$. We are interested in terms linear in the source but an arbitrary order in the SPT terms. We now propose a solution of the form
\beq
\psi_{i}^{ct} =\sum_{n\geq 0} a^{(n+\gamma)} \psi_{i}^{(n+\gamma)},
\eeq
leading to a recursion solution:
\beq
\begin{split}
 \psi_{i,i}^{(n+\gamma)}=& -\frac {1}{g(n+\gamma)-3/2}\left[
\sum_{n_1+n_2=n\atop n_1\geq 0, n_2>0}\left({2 g(n_1+\gamma) - 3/2}\right)\mathcal{L}^{n_1 + \gamma,n_2} + 
\left({2 g(n_2) - 3/2}\right)\mathcal{L}^{n_2 ,n_1+ \gamma} \right.\\
+&\left.\sum_{n_1+n_2+n_3=n\atop n_1\geq 0,n_2,n_3>0}
\left({3 g(n_1+\gamma)-3/2}\right)\mathcal{M}^{n_1+\gamma,n_2,n_3}
+\left({3 g(n_2)-3/2}\right)\left(\mathcal{M}^{n_2,n_1+\gamma,n_3}
+\mathcal{M}^{n_2,n_3,n_1+\gamma}\right)\right. \nonumber \\
+&\left.  S_d^{(n+\gamma)}\right],
\label{eq:rec_count_scalar}
\end{split}
\eeq
where we have introduced the divergence sources $S_d^{(n+\gamma)}$:
\beq
C_{ki} \bar\tau_{i j,kj}= \sum_{n\geq 0} a^{n+\gamma} S_d^{(n+\gamma)}
\eeq

\paragraph*{Vector component of the displacement:}
For the curl part of the displacement field, we start again from the equation of motion (\ref{xfleq}) and multiply it by $A_{ij}$ so as to convert the Eulerian derivative acting on the gravitational potential to a derivative with respect to $q$. We obtain:
\beq
\left( \frac{\partial}{\partial \tau} +\mathcal{H} \right) 
\frac{\partial{x_i}}{\partial q_j} \dot\psi_i
=
\frac{\partial}{\partial q_j} 
\left( \frac{ \dot\psi_i\dot\psi_i}{2} - \phi \right)  - {\cal H}^2  \frac{\partial{x_i}}{\partial q_j}  \frac{\partial \bar\tau^{i k}}{\partial q_k}.
\eeq
We then take curl of that equation to obtain:
\beq
\left( \frac{\partial}{\partial \tau} +\mathcal{H} \right) \epsilon_{lmj} 
\frac{\partial{x_i}}{\partial q_j} \frac{\partial \dot{x}_i}{\partial q_m}
= 
- {\cal H}^2 \epsilon_{lmj}   \frac{\partial{x_i}}{\partial q_j}  \frac{\partial^2 \bar\tau^{i k}}{\partial q_k\partial q_m},
\eeq
so that:
\beq
\left( \frac{\partial}{\partial \tau} +\mathcal{H} \right) \epsilon_{lmj} 
\dot{\psi}_{j,m}
= -\left( \frac{\partial}{\partial \tau} +\mathcal{H} \right) \epsilon_{lmj} 
\psi_{i,j}  \dot{\psi}_{i,m}
- {\cal H}^2 \epsilon_{lmj}   (\delta_{ij} + {\psi_{i,j}} ) {\bar\tau}_{i k,km}.
\eeq
Note that starting from a curl-free initial condition, a curl is generated starting from the third order if $\bar \tau_{ij}$ is absent. We can use this equation to obtain a hierarchy for the curl of $\psi_{i}^{(n+\gamma)}$:
\bea
 \epsilon_{lmj} 
 \psi_{j,m}^{(n+\gamma)}  &=& -\frac{(n+\gamma) + 1/2}{g(n+\gamma) } \sum_{n_1+n_2=n}  \epsilon_{lmj} \left[ (n_2+\gamma) \psi_{i,j}^{(n_1)} \psi_{i,m}^{(n_2+\gamma)}  + n_2  \psi_{i,j}^{(n_1+\gamma)} \psi_{i,m}^{(n_2)} \right] \nonumber \\
 &-& \frac{1}{g(n+\gamma)} {}_cS_l^{(n+\gamma)}
 \eea
 where the curl source ${}_c S$ is defined as
 \beq
 \epsilon_{lmj}   (\delta_{ij} + \psi_{i,j} ) \bar\tau_{i k,km} = \sum_n a^{n+\gamma} {}_{c}S_l^{(n+\gamma)}.
 \eeq
 
 \paragraph*{Lowest order solution:}

We now consider the lowest order solution for the counter terms $\psi_i^{(\gamma)}$. It is convenient to define
\beq
U_i =  \bar\tau_{i k,k}.
\eeq
For the displacement one gets:
\bea\label{divcurl}
\nabla\cdot \psi^{(\gamma)} &=& - \frac{1}{g(\gamma)-3/2}  \nabla\cdot U \nonumber \\
\nabla\times \psi^{(\gamma)} &=& - \frac{1}{g(\gamma)}  \nabla\times U. 
\eea
The relative amplitude of the gradient and curl components depends on the time evolution of the source parametrized by $\gamma$. 
We can write the solution for $\psi^{(\gamma)}$ as:
\bea
\psi^{(\gamma)}&=& - \frac{1}{g(\gamma)-3/2} {\nabla (\nabla\cdot U) \over \nabla^2}  + \frac{1}{g(\gamma)}  {\nabla \times (\nabla\times U) \over \nabla^2}  \nonumber \\
&=& - \frac{3}{2 g(\gamma)(g(\gamma)-3/2)} {\nabla (\nabla\cdot U)\over \nabla^2}  - \frac{1}{g(\gamma)}  U. 
\eea

 \subsection*{Counter Terms}

In the EFT, the stress tensor $\tau^{ij}$ is built using derivatives of the gravitational potential. The equivalence principle requires that there should be at least two spatial derivatives acting on the potential $\d_i\d_j\phi$. Due to the long memory of the system, convective time derivatives should also be included and no price is paid by including an arbitrary number of them. Furthermore one usually writes the stress tensor as an integral in time along the fluid trajectory, again due to the long memory. Because the system is being solved perturbatively, at any given order in initial perturbations $\d_i\d_j\phi$ only has a finite number of terms, each with a specific time dependence. Thus one can choose to write the stress tensor either using a time integral along the fluid trajectory with arbitrary time kernels or a finite set of  convective derivatives. We will choose to write things in terms of convective time derivatives \cite{Mirbabayi2014,Abolhasani}. 

In our current formulation it will be useful to express quantities in Lagrangian space and use derivatives with respect to $q$ rather than $x$. Our goal here is to show that rather than writing the most general counter term using the Eulerian spatial derivatives and convective time derivatives of the gravitational potential we construct the stress tensor simply using $\psi_{i,j}(\vq,\tau)$ and its time derivatives. Furthermore we can trade the time derivative by simply using the $n^{th}$-order perturbative solution of $\psi_{i,j}^{(n)}$ for arbitrary $n$ as the building blocks for our expressions \cite{Mirbabayi2014}.\footnote{Beyond linear order in the counter-terms appropriate combination of convective time derivatives would also isolate $\psi^{(n+\gamma)}_{i,j}$ solutions, and hence they too can be used as building blocks.} 

We can take a spatial derivative of the PT equation:
\beq\label{eqlag1}
\ddot \psi^{i} + \confh \dot \psi^{i} + \partial^i \phi = 0
\eeq
to show that the second derivative of the potential can be written in terms of $\psi_{i,j}$:
\beq
 \partial_i \partial_j \phi = A^{-1}_{k,j} (\ddot \psi_{i,k} + \confh \dot \psi_{i,k}). 
\eeq
Furthermore, convective time derivative of a function of $(\vx,\tau)$ becomes and ordinary time derivative in Lagrangian space. Thus any operator that can be built by combing derivatives of  $\partial_i \partial_j \phi$ can be expressed in terms of $\psi_{i,j}$ and its derivatives. Furthermore given that the perturbative solution of $\psi_{i,j}$ is of the form
\beq
\psi_{i,j} =\sum_n a^n \psi_{i,j}^{(n)}
\eeq
in a Lagrangian perturbative expansion one can use as a basis for all operators $\psi_{i,j}^{(n)}$ instead of $\partial_\tau^n \psi_{i,j}$. To see this define $E_{ij}^{(1)}\equiv \psi_{i,j}$ and note that
\beq
E_{ij}^{(n)} \equiv \confh^{-1}\dot E_{ij}^{(n-1)}- (n-1) E_{ij}^{(n-1)}
\eeq
starts from $n^{th}$ order in the initial perturbations. Furthermore at any fixed order $N$ in perturbation theory one can solve $\{\psi_{i,j}^{(1)},\cdots,\psi_{i,j}^{(N)}\}$ as linear combinations of $\{E_{ij}^{(1)},\cdots,E_{ij}^{(N)}\}$.

Finally when we wrote the equation for $\psi_{i,j}^{(n)}$ we defined $\bar \tau^{ij} \propto C_{il} \tilde \tau^{jl}$. Because $C$ is also just a function of $\psi_{i,j}$ we conclude that the general rule for building all possible counter terms up to a given order is to write the most general tensor $\bar \tau^{ij}$ by contracting any number of $\psi_{k,l}^{(n)}$ in all possible ways such that the final tensor is of the desired order. We will construct all the relevant terms for the two loop power spectrum calculation in a subsequent section. Each of these counter terms provides a specific spatial structure when written in terms of the initial conditions, the amplitude of this term is given by an arbitrary time dependent constant. 
These counter terms source the equation of motion  (\ref{eqlag1}). When solved this equation generates an entire set of spatial structures whose amplitude is calculable given the time dependent amplitude of the original counter term. 

\paragraph*{The need to solve the equation of motion for the counter terms:}

The matrix $C$ is of the form $C_{ij}=\delta_{ij} + C^{(1)}_{ij} + C^{(2)}_{ij} $.  It has a linear ($C^{(1)}_{ij}$) and a quadratic ($C^{(2)}_{ij}$) piece in $\psi_{i,j}$. At lowest order  (\ref{diveq}) becomes: 
\beq\label{eqlag}
\ddot \psi_{i,i}+\confh \dot \psi_{i,i}=\frac{3}{2}\Omega_\text{m}\mathcal{H}^2 \psi_{i,i} - \bar\tau_{i j,ij},
\eeq
where $\bar\tau^{i j}$ is built out of the Lagrangian elements $\psi_{i,j}^{(n)}$. Let us suppose that $U_i= \bar\tau_{i k,k}$ is the gradient of a potential $U_i=\chi_{,i}$ where $\chi$ is a scalar made of $\psi^{(n)}_{i,j}$, and that it is of order $n_\chi$ in perturbations.
In that case the solution of the linear equation is of the form:
\beq
\psi_{i} \propto U_i.
\eeq
Normally, one would plug this back into the interaction terms and includes $C^{(n)}_{ki}\bar\tau_{i j,kj}$ to obtain solutions linear in the counter-term but of order $n_\chi+1$ and higher in initial perturbations. The interaction vertices in equation (\ref{eqlag}) are given by $C^{(1)}_{ki}\psi_{j,k}$ and $C^{(2)}_{ki}\psi_{j,k}$.  However if the $\psi_i$ fields that appear inside $C$ are also gradient (which is true up to $\psi^{(2)}$) then using the explicit form of $C^{(n)}_{ij}$ we get
\beq\label{Cchi}
C^{(n)}_{ki}\chi_{,ik}=  [\chi C^{(n)}_{ki}]_{,ik}.
\eeq
Thus the interaction terms up to order $n_\chi+2$ look like the second derivative of an observable tensor and so they are degenerate with higher order counter terms. Hence to this order one only needs to solve the equation of motion for the counter terms for which $U_i$ has a curl component. Most of the stress tensor counter terms that appear at the lowest orders satisfy this simplifying assumption. 
The only place where  the time dependence of the counter terms is relevant is when solving the equations of motion to obtain the higher order contributions. Although the time dependence also affects the linear solution, that dependency can be absorbed into the overall normalization. In other words given that the time dependence of the counter terms in the EFT is treated as unknown, so long as the result of integrating the equations of motion to higher orders is degenerate with other higher order counter terms one can trade one unknown function of time at the level of the equation of motion with one free function multiplying the final solution. One can pick any arbitrary time dependence and solve the equations using it. The difference in the result among different choices can be reabsorbed into the higher order counter terms. It will prove useful for the linear counter term to choose a time dependence for the source such that the linear solution it leads to has the same time dependence as the linear theory solution in SPT.

\subsection*{Examples:}

As an illustration we will now list the two loop counter terms with the goal of understanding their contribution to the two-loop power spectrum. 

\subsubsection{Linear counter terms}

The linear displacement is the gradient of a potential $\psi_k^{(1)}(\vq)=\phi^{(1)}_{,k}(\vq)$, where $\phi^{(1)}(\vq)=\Delta_{\vq}^{-1} \delta_0(\vq)$ is the initial condition for the gravitational potential multiplied by the growth factor of the final time. In what follows we only consider quantities at a single time, drop the time arguments, and absorb all growth factors into the initial Gaussian fluctuations. 

At the lowest order there are two possible counter terms:
\bea
\bar\tau_{ij}^{(1)} &\propto& \delta_{ij} \psi^{(1)}_{k,k}=\delta_{ij} \phi_{,kk}^{(1)} \nonumber \\
\bar\tau_{ij}^{(1)} &\propto&  \psi^{(1)}_{i,j}=\phi_{,ij}^{(1)}.
\eea
In both cases 
\beq
U_i=\bar\tau_{ij,j}\propto (\Delta \phi^{(1)})_{,i}
\eeq
Thus up to third order in $\delta_0$ the terms that are generated by solving the non-linear equations of motion with this source term are degenerate with higher order counter terms. For two-loop power spectrum the counter terms have to be included in tree-level and one-loop diagrams, i.e. one needs up to the third order solution in $\delta_0$. Hence we do not need to solve the associated equations of motion and the time dependence of these sources can be absorbed into the amplitude of higher order counter terms. Thus this source only adds to the perturbation theory solution the term
\beq\label{ct11}
\psi_k^{ct (1)} = a_1 (\Delta\phi^{(1)})_{,k}.
\eeq

\subsubsection{Quadratic counter terms}

At this order we can construct counter terms by either using the gradient of the second order displacement, $\psi_{k,l}^{(2)}(\vq)$ or two first order ones $\psi_{k,l}^{(1)}(\vq)\psi_{m,n}^{(1)}(\vq)$. The second order displacement is also the gradient of a potential, 
$\psi_k^{(2)}(\vq)=\phi^{(2)}_{,k}(\vq)$. The general structure of the perturbation theory equation  (\ref{diveqpert}) is such that at any order the expression for $\psi^{(n)}_{k,k}$ can be solved for in terms of local expressions containing lower order $\psi^{(m)}_{i,j}$ with $m<n$. The counter terms that can be constructed out of $\psi_k^{(2)}(\vq)$,
\beq
\bar\tau_{ij,j}^{(2)} \propto \phi^{(2)}_{,ijj}= \psi_{j,ji}^{(2)}
\eeq
can thus be rewritten in terms of quadratic combinations of $\psi_{i,j}^{(1)}$, and they do not need to be considered separately. 

There are three possible combinations of two $\psi_{i,j}^{(1)}=\phi_{,ij}^{(1)}$ to consider:
\beq
\bar\tau_{ij}^{(2)} \propto b_1 \delta_{ij} (\phi_{,kk}^{(1)})^2  + b_2  \delta_{ij} \phi_{,kl}^{(1)}\phi_{,kl}^{(1)} + b_3  \phi_{,ik}^{(1)}\phi_{,kj}^{(1)} 
\eeq
We can compute 
\beq\label{quadct}
\bar\tau_{il,lj}^{(2)} \propto
 b_1  (\phi_{,kk}^{(1)})_{,ij}^2   +  b_2  (\phi_{,kl}^{(1)}\phi_{,kl}^{(1)})_{,ij} + b_3  (\phi_{,ik}^{(1)}\phi_{,kl}^{(1)})_{,lj}.  
\eeq
For the terms proportional to $b_1$ and $b_2$, $U_i$ is a pure gradient and thus for those we do not need to solve the equations of motion and their time dependence is irrelevant. We could have also written:
\beq
\bar\tau_{ij}^{(2)} \propto b_4  \phi_{,ij}^{(1)} \phi_{,kk}^{(1)}, 
\eeq
however 
\beq
(\phi_{,ij}^{(1)} \phi_{,kk}^{(1)})_{,j}=\left( \frac{1}{2}\delta^{ij} \left[(\phi_{,kk}^{(1)})^2 - \phi_{,kl}^{(1)}\phi_{,kl}^{(1)} \right]+ \phi_{,ik}^{(1)}\phi_{,kj}^{(1)} \right)_{,j},
\eeq
and thus this term is degenerate with what we already have. 

\subsubsection{Cubic counter terms}

At this order we can construct counter terms in three different ways. We can combine three first order terms, 
$\psi_{k,l}^{(1)}(\vq)$, one first order and one second order term $\psi_{k,l}^{(1)}(\vq)\psi_{l,m}^{(2)}(\vq)$ or one third order term $\psi_{k,l}^{(3)}(\vq)$. 
We get:
\bea\label{ct3}
\bar\tau_{ij}^{(3)} &\propto&  c_1 \delta_{ij} (\phi_{,kk}^{(1)})^3  + c_2  \phi_{,ik}^{(1)}\phi_{,kj}^{(1)}\phi_{,mm}^{(1)} + c_3  \delta_{ij} \phi_{,kl}^{(1)}\phi_{,kl}^{(1)} \phi_{,mm}^{(1)} + c_4  \phi_{,ik}^{(1)}\phi_{,kl}^{(1)}\phi_{,lj}^{(1)} \nonumber\\[8pt]
&+& c_5  \delta_{ij} \phi_{,kl}^{(1)}\phi_{,lm}^{(1)} \phi_{,mk}^{(1)} + c_6 \phi_{,kl}^{(1)}\phi_{,kl}^{(1)} \phi_{,ij}^{(1)}
+c_7 \phi_{,ij}^{(1)}\phi_{,mm}^{(1)} \phi_{,nn}^{(1)}\nonumber \\[8pt]
&+&{c_8 \over 2}  (\phi_{,ik}^{(1)}\phi_{,kj}^{(2)}+\phi_{,ik}^{(2)}\phi_{,kj}^{(1)}) + c_9  \delta_{ij} \phi_{,kl}^{(1)}\phi_{,kl}^{(2)} \nonumber \\[8pt]
&+&  c_{10} (\psi_{i,j}^{(3)} + \psi_{j,i}^{(3)})
\eea
We have not included any term proportional to $\psi_{k,k}^{(2)}$ or $\psi_{k,k}^{(3)}$ as they are degenerate with the terms constructed out of lower order displacement fields. 

\section{Eulerian density}

The goal of this section is to obtain an expression for the density field and compute the corrections to the power spectrum coming from all the counter terms, and to prove equation (\ref{summary}).

So far we have solved the Eulerian EFT equations by solving for $\vps(\vq,\tau)$ such that $\vx(\vq,\tau)= \vq + \vps(\vq,\tau)$ with $\vps(\vq,0)=0$. We are interested in computing the density field which is obtained using the standard formulas of Lagrangian perturbation theory.  

We can obtain the final density directly from $J$ which we have solved explicitly in terms of $\vq$. Using \eqref{measure}
\beq
\delta (\vx) = \Big({1\over J(\vq)} -1\Big)_{\vx=\vq+\vps} 
= \Big({1\over \left(1+\mathcal{K}+\mathcal{L}+\mathcal{M}\right)}-1\Big)_{\vx=\vq+\vps}
\eeq
This expression should be used perturbatively expanding up to a given order in the displacement field. 
In reality $J$ depends only on $\psi_{i,j}$ while there are non-linear terms from the mapping that depend on $\psi_i$. These two sets of terms can have different sizes and the latter should be resummed at the BAO scale. 

One can also obtain an equation in Fourier space
\beq\label{deltaF}
\delta(\vk)=\int d^3 q \ \ e^{i \vk\cdot (\vq+\vps)},
\eeq
again to be treated perturbatively. We can take a time derivative of this equation and obtain:
\beq
\dot\delta(\vk)=i \int d^3 q \ \vk\cdot\dot\vps \ e^{i \vk\cdot (\vq+\vps)} = i \vk\cdot\int d^3 x \ (1+\delta) \vv \ e^{i \vk\cdot \vx}.
\eeq
Thus in this formulation the continuity equation does not have any additional terms. This implies that it is the density and momentum that are being treated as primary variables and the velocity (and $\psi$) as composite operators (see e.g. \cite{Abolhasani} for a discussion on this point). 

These formulas are identical to those of standard Lagrangian perturbation theory except for the additional contribution to the displacement field coming from the counter terms. These equations, if expanded consistently to a given order in the power spectrum, give the same answer as the standard SPT formulas. Thus all we need to do is to keep track of the contributions from the counter terms. For the two loop power spectrum calculation we are only interested in contributions linear in the counter terms.  The lowest order counter term counts as third order in perturbation theory and thus its squared is a sixth order contribution to the power spectrum which is the same as two-loop but is simply given by $k^4 l^4 P(k)$. The coefficient of this term is however a free parameter as one can directly write $\bar \tau_{ij,ij} = l^4 \Delta^2 \delta$ as a new linear, higher derivative, counter term. 

Expanding equation (\ref{deltaF}) up to linear order in the counter terms and expressing everything in Fourier space we get:
\beq\label{deltaF2}
\delta(\vk)=\delta^{SPT}(\vk)+ \sum_n {i^n \over (n-1)!} \int {d^3 p_1 \over (2\pi)^3} \cdots {d^3 p_n \over (2\pi)^3} (2\pi)^3 \delta^D(\vp_1+\cdots \vp_n - \vk) \vk\cdot \vps^{ct}(\vp_1) \vk\cdot\vps^{SPT}(\vp_2) \cdots \vk\cdot\vps^{SPT}(\vp_n)
\eeq
where $\delta^D$ indicates Dirac delta function. In this expression, the SPT solutions need to be expanded and only terms up to a fixed order kept. 
To simplify our notation from now on we will use
\beq
\int {d^3 p_1 \over (2\pi)^3} \cdots {d^3 p_n \over (2\pi)^3} (2\pi)^3 \delta^D(\vp_1+\cdots \vp_n - \vk)\rightarrow \int_p
\eeq

We also found that for all terms, except one of the second order counter terms, the terms generated by solving the equations of motion are degenerate with higher order counter terms. Thus  for these terms, we could decide not to solve those equations of motion of SPT, or solve them assuming any particular time dependence. An additional simplification happens if we assume that the linear solution sourced by the counter terms has the same time dependence as the linear theory solution. If that was the case, then there is a very simple way to obtain an expression for the density; one can simply replace the initial density perturbation $\delta_0(\vp)$ in the usual SPT solutions for $\delta_0(\vp) \rightarrow \delta_0(\vp) + \delta_0^{ct}(\vp)$ and keep up to the linear terms in the counter terms. In this way one can easily recycle the usual SPT Kernels. In SPT
\beq
\delta^{SPT}(\vk)=\sum_{n\geq 1} \int_p  F_n(\vp_1,\cdots,\vp_n) \delta_0(\vp_1) \cdots  \delta_0(\vp_n).
\eeq
Now the solution associated with any given counter term becomes:
\beq\label{deltact}
\delta^{ct}(\vk)= \sum_{n\geq 1} n \int_p F_n(\vp_1,\cdots,\vp_n) \delta_0^{ct}(\vp_1)\delta_0(\vp_2) \cdots  \delta_0(\vp_n).
\eeq
where we used the fact that $F_n$ is symmetric under permutation of its arguments.

\subsection{Linear counter terms}

The lowest order counter term corresponds to choosing \eqref{ct11}
\beq\label{ct1}
\vps^{ct1(1)}(\vk) = -i l_1^2 \vk \delta_0(\vk).
\eeq
Note that as argued below \eqref{Cchi} up to third order in $\delta_0$, relevant for the 2-loop power spectrum, higher order solutions $\vps^{ct1(2)},\vps^{ct1(3)}$ are degenerate with higher order counter terms. To compute all the terms associated with this terms we will need the linear and quadratic SPT solutions:
\bea\label{ct2}
\vps^{SPT(1)}(\vk) &=& - {i \vk \over \vk^2} \delta_0(\vk) \nonumber \\
\vps^{SPT(2)}(\vk) &=& - {3\over 14}  {i \vk \over \vk^2} \int_p \left(1- {(\vp_1\cdot\vp_2)^2  \over \vp_1^2 \vp_2^2}\right) \delta_0(\vp_1)\delta_0(\vp_2) 
\eea
We now use equation (\ref{deltaF2}) to obtain all the terms associated with the linear counter term:
\bea
\delta^{ct1}(\vk)&=& i \vk\cdot \vps^{ct(1)}(\vk) \nonumber \\
&+& i^2 \int_p  \vk\cdot \vps^{ct(1)}(\vp_1)\ \vk\cdot \vps^{SPT(1)}(\vp_2)\nonumber \\
&+&  i^2 \int_p  \vk\cdot \vps^{ct(1)}(\vp_1) \ \vk\cdot \vps^{SPT(2)}(\vp_2) + {1\over 2}  i^3 \int_p \vk\cdot \vps^{ct(1)}(\vp_1)\  \vk\cdot \vps^{SPT(1)}(\vp_2)\ \vk\cdot \vps^{SPT(1)}(\vp_3)
\eea
Using equations (\ref{ct1}) and (\ref{ct2}) we get
\bea\label{ct12}
\delta^{ct1}(\vk)&=& l_1^2 \vk^2 \delta_0(\vk) \nonumber \\
&+& l_1^2 \int_p  {\vk\cdot\vp_1 \over \vp_1^2} {\vk\cdot\vp_2 \over \vp_2^2} \vp_1^2  \delta_0(\vp_1)\delta_0(\vp_2) \nonumber \\
&+&  l_1^2  \int_p  {\vk\cdot\vp_1 \over \vp_1^2}\left[ {1\over 2} {\vk\cdot\vp_{2} \over \vp_{2}^2} {\vk\cdot\vp_{3} \over \vp_{3}^2} + {3 \over 14} {\vk\cdot\vp_{23} \over \vp_{23}^2} \left(1- {(\vp_2\cdot\vp_3)^2  \over \vp_2^2 \vp_3^2}\right) \right] \vp_1^2  \delta_0(\vp_1)\delta_0(\vp_2)\delta_0(\vp_3)
\eea
We can obtain an equally valid expression starting from equation (\ref{deltact}):
\bea
\bar \delta^{ct(1)}(\vk)&=& l_1^2 \sum_n n \int_p F_n(\vp_1,\cdots,\vp_n) \vp_1^2 \delta_0(\vp_1)\delta_0(\vp_2) \cdots  \delta_0(\vp_n).
\eea
This implies that to the order relevant for 2-loop power spectrum we have
\beq\label{ct1Fn}
\langle \delta^{ct1}\delta^{SPT}\rangle = 
\frac{1}{2}\Big(\langle \delta^{SPT}\delta^{SPT}\rangle_{P(q)\to (1+l_1^2 q^2)^2P(q)}
-\langle \delta^{SPT}\delta^{SPT}\rangle\Big) +\mathcal{O}(l_1^4),
\eeq
where in the first term on the right-hand side all linear power spectra in the SPT calculation must be replaced as indicated. The difference between this and using \eqref{ct12} is degenerate with quadratic and higher order counter terms. The result is 
\bea
\langle \delta^{ct1}\delta^{SPT}\rangle &=& l_1^2 \vk^2 (P(k)+P_{13}(k)) \nonumber \\
&+& 2 l_1^2 \int_p  {\vk\cdot\vp_1 \over \vp_1^2} {\vk\cdot\vp_2 \over \vp_2^2} \vp_2^2 F_2(\vp_1,\vp_2)P(p_1)P(p_2)\nonumber \\
&+& l_1^2  \int_p  {\vk\cdot\vp_1 \over \vp_1^2}\left[ {1\over 2} {\vk\cdot\vp_{2} \over \vp_{2}^2} {\vk\cdot\vp_{3} \over \vp_{3}^2} + {3 \over 14} {\vk\cdot\vp_{23} \over \vp_{23}^2} \left(1- {(\vp_2\cdot\vp_3)^2  \over \vp_2^2 \vp_3^2}\right)\right]
\nonumber\\
&&~~~~~~~~~~~~~~~\left[2P(p_1)P(k) (2\pi)^3\delta^D(\vp_1+\vp_2)+P(p_1)P(p_2) (2\pi)^3\delta^D(\vp_2+\vp_3)\right] 
\eea
where 
\bea
F_2(\vp_1,\vp_2) &=& {5 \over 7} + {2 \over 7} {(\vp_1\cdot\vp_2)^2  \over \vp_1^2 \vp_2^2} + {1\over 2}  \left({\vp_1\cdot\vp_2\
\over \vp_1^2}+ {\vp_1\cdot\vp_2\over \vp_2^2}\right) \nonumber \\
&=&   {3 \over 14} \left(1- {(\vp_1\cdot\vp_2)^2  \over \vp_1^2 \vp_2^2}\right) + {1 \over 2} {\vp_{12}\cdot\vp_1\over \vp_1^2} 
{\vp_{12}\cdot\vp_2\over \vp_2^2 },
\eea
and $P_{13}(k) = \langle \delta^{SPT(3)}\delta^{SPT(1)}\rangle$. Now let us define $\vp_1\cdot\vk=k p_1 x$ and $r=p_1/k$. It is also useful to define $y(r, x)= 1+r^2-2 r x$. With these definitions,
\beq\label{F2}
F_2(\vk-\vp,\vp)= {3 \over 14} {(1-x^2)\over y(r,x)} +  {1 \over 2} {(1-r x) x\over r y(r,x)},
\eeq 
We then get 
\bea
\langle \delta^{ct1} \delta^{SPT} \rangle &=&  l_1^2 k^2 (P(k)+P_{13}(k)) \nonumber \\
&+& 2 l_1^2k^5 \int {dr r^2 \over (2 \pi)^2 } P(k r) \int_{-1}^1 dx P(k \sqrt{y(r,x)}) \left[{3 \over 14} {(1-x^2)\over y(r,x)} 
+ {1 \over 2} {(1-r x) x\over r y(r,x)}\right] \frac{x(1-rx)}{r}\nonumber \\
&+&   2 l_1^2k^5 P(k)\int {dr r^2 \over (2 \pi)^2 } P(k r) 
 \int_{-1}^1 dx \left[-\frac{1}{2}x^2(1+\frac{1}{2r^2})+\frac{3}{14}\frac{rx(1-rx)(1-x^2)}{y(r,x)}\right].
\eea

Note that ignoring two loop contributions of the form $k^2 P(k)$ corresponds to adding a correction $\Delta l_1^2$ which cancels them. The value of $l_1^2$ which appears in the above formula is the one which is fixed at the one loop order. $\Delta l_1^2$ has to be taken into account at three and higher loops. Note also that in the limit where one of the loop momenta goes to zero (i.e. $r\to 0$) $P_{13}(k)$ gives 
\beq
-\frac{1}{2}P(k)\int {dr r^2 \over (2 \pi)^2 } P(k r)  \int_{-1}^1 dx \frac{x^2}{r^2},
\eeq
so that the infra-red singularities cancel among various terms.

Starting from equation (\ref{ct1Fn}) we can obtain an equivalent expression for the power spectrum:
\bea
\langle \delta^{ct1} \delta^{SPT} \rangle &=&  l_1^2 k^2 \left(P(k)+P_{22}(k)+2 P_{13}(k)+{\bar P}_{22}(k)+2 {\bar P}_{13}(k)\right), 
\eea
where
\bea
{P}_{22}(k)&=& \int_p  2 F_2^2(\vp,\vk-\vp) P(p) P(|\vk-\vp|)\nonumber \\
{P}_{13}(k)&=& P(k) \int_p 3 F_3(\vk,\vp,-\vp)  P(p) \nonumber \\
{\bar P}_{22}(k)&=& \int_p 4 F_2^2(\vp,\vk-\vp) {\vp\cdot(\vp-\vk)\over k^2} P(p) P(|\vk-\vp|) \nonumber \\
{\bar P}_{13}(k)&=& P(k) \int_p 3 F_3(\vk,\vp,-\vp) {p^2 \over k^2}  P(p),
\eea

\subsection{Quadratic counter terms}

The lowest order solution sourced by the second order counter terms can be written as: 
\bea\label{quadsol}
\vps^{ct2(2)}(\vp_{12}) &=& - {i \vp_{12} \over \vp_{12}^2} \left( \int_p [l_{21}^2 \vp_{12}^2  +  l_{22}^2 \vp_{12}^2 (1- {  (\vp_1\cdot\vp_2)^2 \over \vp_1^2 \vp_2^2} )+ l_{23}^2  {\vp_{12}\cdot\vp_1\  \vp_{12}\cdot\vp_2 \ \vp_1\cdot\vp_2 \over \vp_1^2 \vp_2^2} ] \delta_0(\vp_1)\delta_0(\vp_2) \right) \nonumber \\
&+& i {\bar l_{23}^2 }\  \int_p {\vp_{12}\times(\vp_{12}\times\vp_1)\ (\vp_{12}\cdot\vp_2)  \ \vp_1\cdot\vp_2 \over \vp_1^2 \vp_2^2 \vp_{12}^2} \delta_0(\vp_1)\delta_0(\vp_2) \nonumber \\
&=& - {i \vp_{12} \over \vp_{12}^2} \left( \int_p [l_{21}^2 \vp_{12}^2  +  l_{22}^2 \vp_{12}^2 (1- {  (\vp_1\cdot\vp_2)^2 \over \vp_1^2 \vp_2^2} )+ (l_{23}^2 -  \bar l_{23}^2) {\vp_{12}\cdot\vp_1\  \vp_{12}\cdot\vp_2 \ \vp_1\cdot\vp_2 \over \vp_1^2 \vp_2^2} ] \delta_0(\vp_1)\delta_0(\vp_2) \right) \nonumber \\
&-& i {\bar l_{23}^2 }\  \int_p {\vp_1\ (\vp_{12}\cdot\vp_2)  \ \vp_1\cdot\vp_2 \over \vp_1^2 \vp_2^2 } \delta_0(\vp_1)\delta_0(\vp_2) 
\eea
where $l_{22}$ is a linear combination of $b_1$ and $b_2$ in \eqref{quadct}. Here we see that the term proportional to  $l_{22}^2$ can be used to cancel the terms that depended on the assumed time dependence of the linear counter term. The terms proportional to $l_{23}^2$ and $\bar l_{23}^2$ originate from the same stress tensor counter terms but the ratio of their amplitudes depends on the time dependence of this component, $(g(\gamma) -3/2) l_{23}^2 = g(\gamma) \bar l_{23}^2$, where $\gamma$ parametrizes the time dependence of the counter term (see \eqref{divcurl}). 

At the order we are interested, we should use equation (\ref{deltaF2}) and combine these quadratic counter terms to produce up to third order terms: 
\bea
\delta^{ct2}(\vk)&=& i \vk\cdot (\vps^{ct2(2)}(\vk)+\vps^{ct2(3)}(\vk)) \nonumber \\
&+& i^2 \int_p  \vk\cdot \vps^{ct2(2)}(\vp_1)\ \vk\cdot \vps^{SPT(1)}(\vp_2)
\eea
where $\vps^{ct2(3)}(\vk)$ is the third order displacement obtained from solving the equations of motion in the presence of the only quadratic source which is not curl-free, namely $b_3$ in \eqref{quadct}.

\paragraph*{Quadratic Terms:} The quadratic terms that derive from the quadratic counter terms $\delta^{ct2(2)}(\vk)$ then become:
\beq
\delta^{ct2(2)}(\vk)= \int_p  \left[ l_{21}^2 \vp_{12}^2 +  l_{22}^2  \vp_{12}^2 (1- {   (\vp_1\cdot\vp_2)^2 \over \vp_1^2 \vp_2^2}) + l_{23}^2   {\vp_{12}\cdot\vp_1\  \vp_{12}\cdot\vp_2 \ \vp_1\cdot\vp_2 \over \vp_1^2 \vp_2^2}\right]   \delta_0(\vp_1)\delta_0(\vp_2) 
\eeq
where we have defined $\vk=\vp_{12}=\vp_1+\vp_2$. 
The  contribution to the power spectrum is
\beq
\langle \delta^{ct2(2)}\delta^{SPT(2)}\rangle=2  \int_p \left[ l_{21}^2  \vk^2 + l_{22}^2 \vk^2 (1-{  (\vp_1\cdot\vp_2)^2 \over \vp_1^2 \vp_2^2} )+ l_{23}^2  {\vk\cdot\vp_1\  \vk\cdot\vp_2 \ \vp_1\cdot\vp_2 \over \vp_1^2 \vp_2^2} \right]F_2(\vp_1,\vp_2) P(p_1)P(p_2)
\eeq
We now write the integrals one needs to do to compute the two-loop counter terms. Using the same definitions given above \eqref{F2} we get 
\bea
\langle \delta^{ct2(2)}\delta^{SPT(2)}\rangle &=& 2 k^3 \int {dr r^2 \over (2 \pi)^2 } P(k r) \int_{-1}^1 dx P(k \sqrt{y(r,x)}) \left[{3 \over 14} {(1-x^2)\over y(r,x)} +  {1 \over 2} {(1-r x) x\over r y(r,x)}\right] \nonumber \\
&\times& \left[ k^2 l_{21}^2  +  k^2  { l}_{22}^2   {(1-x^2)\over y(r,x)} + k^2  {l}_{23}^2 {(1-r x) (x-r) x\over  y(r,x)}  \right ]. 
\eea

\paragraph*{Cubic Terms from the mapping between displacement and density:} We now discuss the cubic pieces that derive from the quadratic counter term:
\bea
 \delta^{ct2(3a)}(\vk)&=& i^2 \int_p  \vk\cdot \vps^{ct(2)}(\vp_{12})\ \vk\cdot \vps^{SPT(1)}(\vp_3) \nonumber \\
 &=&  \int_p \left[ {\vk\cdot \vp_3 \over \vp_3^2} {\vk\cdot\vp_{12}\over \vp_{12}^2} \left( l_{21}^2 \vp_{12}^2 +  l_{22}^2  \vp_{12}^2 (1- {   (\vp_1\cdot\vp_2)^2 \over \vp_1^2 \vp_2^2}) + l_{23}^2   {\vp_{12}\cdot\vp_1\  \vp_{12}\cdot\vp_2 \ \vp_1\cdot\vp_2 \over \vp_1^2 \vp_2^2}\right) \right. \nonumber \\
 &-& \left. {\bar l_{23}^2 } {\vk \cdot \vp_3 \ \vk\cdot(\vp_{12}\times(\vp_{12}\times\vp_1))\ (\vp_{12}\cdot\vp_2)  \ \vp_1\cdot\vp_2 \over \vp_1^2 \vp_2^2 \vp_3^2 \vp_{12}^2}\right] \nonumber \\
 &\times& \delta_0(\vp_1)\delta_0(\vp_2) \delta_0(\vp_3). 
\eea

For the contribution to the power spectrum we get
\bea
\langle \delta^{ct2(3)}\delta^{SPT(1)}\rangle&=&2  \int_p  
\left[ {\vk\cdot \vp_3 \over \vp_3^2} {\vk\cdot\vp_{12}\over \vp_{12}^2} \left( l_{21}^2 \vp_{12}^2 +  l_{22}^2  \vp_{12}^2 (1- {   (\vp_1\cdot\vp_2)^2 \over \vp_1^2 \vp_2^2}) + l_{23}^2   {\vp_{12}\cdot\vp_1\  \vp_{12}\cdot\vp_2 \ \vp_1\cdot\vp_2 \over \vp_1^2 \vp_2^2}\right) \right. \nonumber \\
 &-& \left. \frac{\bar l_{23}^2 }{2} {\vk\cdot\vp_3 \ 
\vk\cdot (\vp_{12}\times(\vp_{12}\times\vp_1))\ (\vp_{12}\cdot\vp_2)  \ \vp_1\cdot\vp_2 
\over \vp_1^2 \vp_2^2 \vp_3^2 \vp_{12}^2}\right] \nonumber \\
&\times& (2\pi)^3 \delta^D(\vp_2+\vp_3) (2\pi)^2 \delta^D(\vp_1-\vk) P(k)P(p_3). 
\eea

We can define $\vp_3\cdot\vk = p_3 k x$ and $r=p_3/k$ and write
\bea
\langle \delta^{ct2(3a)}\delta^{SPT(1)}\rangle &=& 2 k^3 \int {dr r^2 \over (2 \pi)^2 } P(k r) P(k) \int_{-1}^1 dx   \nonumber \\
&\times& \left( \left[{x (1-r x) \over r y(r,x)}\right] \left[ k^2 l_{21}^2  y(r,x) +  k^2  { l}_{22}^2   {(1-x^2) y(r,x)} - k^2  {l}_{23}^2 {(1-r x) (r-x) x}  \right ] \right. \nonumber \\
&-& \left. {\bar {l}_{23}^2 \over 2} {x^2 (1-x^2) (r^2-1) \over y} \right). 
\eea

In the terms proportional $l_{21}^2$ and $l_{22}^2$ the $y$ denominator cancels and thus the result is proportional to $k^2 P$:
\beq
\langle \delta^{ct2(3a)}\delta^{SPT(1)}\rangle _{(l_{21}^2,l_{22}^2)} = \left[ {2\over 3} k^2 l_{21}^2 +{4\over 15} l_{22}^2\right] P(k) \int {d^3\vp \over (2\pi)^3} P(p).
\eeq
These terms are proportional to counter terms we already have so they are not relevant for the power spectrum calculation:
The terms proportional to $l_{23}^2$ and $\bar l_{23}^2$ are the non-trivial ones. For the first one we get
\bea
\langle \delta^{ct2(3a)}\delta^{SPT(1)}\rangle _{l_{23}^2} &=&- k^2  {l}_{23}^2 P(k) \nonumber \\
&\times& 2 k^3 \int {dr r^2 \over (2 \pi)^2 } P(k r)  \int_{-1}^1 dx  \left[{x^2 (1-r x)^2 (r-x) \over r y(r,x)}\right] \nonumber \\
&=&- k^2  {l}_{23}^2 P(k) \nonumber \\
&\times& 2 k^3 \int {dr r^2 \over (2 \pi)^2 } P(k r) ({2\over 5} +  \int_{-1}^1 dx  \left[{x^2 (1-r x) (1-x^2) \over y(r,x)}\right] ),
\eea
where we have used the fact that $(1-rx)(r-x)=r(1-x^2)- x y(r,x)$. For the second one we have:
\bea
\langle \delta^{ct2(3a)}\delta^{SPT(1)}\rangle _{\bar l_{23}^2} &=& - k^2 { {\bar l}_{23}^2 \over 2} P(k) \nonumber \\
&\times& 2 k^3 \int {dr r^2 \over (2 \pi)^2 } P(k r)  \int_{-1}^1 dx  \left[{x^2 (1-x^2) (r^2-1) \over  y(r,x)}\right] \nonumber \\
&=& k^2  {\bar l}_{23}^2 P(k) \nonumber \\
&\times& 2 k^3 \int {dr r^2 \over (2 \pi)^2 } P(k r) (-{4\over 30} +  \int_{-1}^1 dx  \left[{x^2 (1-r x) (1-x^2) \over y(r,x)}\right] ),
\eea
where we have used that $r^2-1 = y - 2(1-r x)$.  Thus both contributions, the contributions proportional to $l_{23}^2$ and $\bar l_{23}^2$ have the same $k$ dependence. This could have easily been derived using the second form of the solution in equation (\ref{quadsol}). The contribution to $\vps^{ct(2)}$ that has no $\vp_{12}^2$ in the denominator clearly leads to a contribution proportional to $k^2 P$ which we can thus absorb into an existing counter terms.

\paragraph*{Cubic Terms from the equations of motion:} Finally let us discuss the contribution coming from the equations of motion when sourced by the counter term proportional to $l_{23}^2$ and $\bar l_{23}^2$. We have already shown that the other two terms lead to $k$-dependences that are degenerate with higher order counter terms. Hence, for the relevant part of the source we can write
\beq
U_i = \bar\tau_{ij,j} = - (g(\gamma)-3/2)\psi^{ct23(2)}_i - g(\gamma)\psi_i^{ct \bar {23}(2)}
\eeq
where $\psi^{ct23(2)}_i$ and $\psi^{ct\bar{23}(2)}_i$ refer respectively to the pieces proportional to $l_{23}^2$ and $\bar l_{23}^2$ in \eqref{quadsol}. For the $l_{23}^2$ we get:
\bea
 \delta^{ct2(3b)}(\vk)
 &=& h_1(\gamma)  l_{23}^2 \int_p {\vp_3^2  \vp_{12}^2- (\vp_3\cdot\vp_{12})^2\over \vp_3^2  \vp_{12}^2}  \left[  {\vp_{12}\cdot\vp_1\  \vp_{12}\cdot\vp_2 \ \vp_1\cdot\vp_2 \over \vp_1^2 \vp_2^2}\right] \nonumber \\
 &\times& \delta_0(\vp_1)\delta_0(\vp_2) \delta_0(\vp_3), 
\eea
where $h_1(\gamma)=-g(1)/(g(1+\gamma)-3/2)$ is a constant that depends on the time dependence of this counter term. 

Although the momentum structure in this term is not degenerate with what we obtain for higher order counter terms, if we are interested in computing the power spectrum, then the contribution will be identical to one of the terms we will consider in the next section. When computing the power spectrum, the $\delta_0(\vp_3)$ will have to be contracted with either $\delta_0(\vp_1)$ or $\delta_0(\vp_2)$ both giving the same answer.  Thus we can set $\vp_3=-\vp_2$ and use the fact that $\vp_{123}=\vp_1=\vk$. Thus we can replace:
\bea
\vp_3^2  \vp_{12}^2- (\vp_3\cdot\vp_{12})^2&\rightarrow& \vp_1^2  \vp_{2}^2- (\vp_1\cdot\vp_{2})^2 \nonumber \\
\vp_{12}\cdot\vp_1\  \vp_{12}\cdot\vp_2 \ \vp_1\cdot\vp_2 &\rightarrow& \vp_{12}\cdot\vp_{123}\  \vp_{12}\cdot\vp_3 \ \vp_{123}\cdot\vp_3.
\eea 
After this replacements we get:
\bea\label{eqm13}
 \delta^{ct2(3b)}(\vk)
 &\rightarrow& h_1(\gamma)  l_{23}^2  \int_p {\vp_1^2  \vp_{2}^2- (\vp_1\cdot\vp_{2})^2\over \vp_2^2  \vp_{12}^2}  \left[  {\vp_{12}\cdot\vp_{123}\  \vp_{12}\cdot\vp_3 \ \vp_{123}\cdot\vp_3 \over \vp_1^2 \vp_3^2}\right] \nonumber \\
 &\times& \delta_0(\vp_1)\delta_0(\vp_2) \delta_0(\vp_3). 
\eea
This is the same momentum structure of one of the cubic counter terms, namely $c_8$ in \eqref{ct3}.

Finally let us look at the term arising from $\bar l_{23}^2$, we get:
\bea
 \delta^{ct2(3b)}(\vk)
 &=& - h_2(\gamma)  {\bar l_{23}^2\over 2}  \int_p { \vp_3\cdot\vp_{12}  (\vp_3\cdot(\vp_{12}\times(\vp_1\times\vp_2))\ (\vp_2^2-\vp_1^2)  \  \vp_1\cdot\vp_2 \over \vp_1^2 \vp_2^2 \vp_3^2 \vp_{12}^2} \nonumber \\
 &\times& \delta_0(\vp_1)\delta_0(\vp_2) \delta_0(\vp_3), 
\eea
where $h_2(\gamma)=(g(1)-3/2)/(g(1+\gamma)-3/2)$ is a constant that depends on the time dependence of this counter term. This is a different momentum dependence, but if we are interested in computing the power spectrum, we can  interchange $\vp_3=-\vp_2$ and use the fact that $\vp_{123}=\vp_1$. Thus we can replace:
\bea
(\vp_3\cdot(\vp_{12}\times(\vp_1\times\vp_2))&\rightarrow& \vp_1^2  \vp_{2}^2- (\vp_1\cdot\vp_{2})^2 \nonumber \\
(\vp_2^2-\vp_1^2) &\rightarrow& \vp_{12}^2 - 2 \vp_{12}\cdot\vp_{123}  \nonumber \\
\vp_1\cdot\vp_2 &\rightarrow& -\vp_{123}\cdot\vp_3.
\eea 
Thus after integrating over $\vp_2$ the momentum dependence is identical to that in equation (\ref{eqm13}) apart from a term proportional to $k^2 P$. This again should be clear from the second form of the solution in equation (\ref{quadsol}). Thus the momentum dependence of this contribution to the power spectrum is also degenerate with one of the higher order counter terms.

\subsection{Cubic counter terms}

At the order we are working we only need the linear contribution from the cubic displacement to the density:
\beq
\delta^{ct3}(\vk)= i \vk\cdot \vps^{ct3(3)}(\vk)
\eeq
The terms proportional to $c_1-c_7$ in equation (\ref{ct3}) lead to the following contributions: 
\bea
\delta^{ct3}|_{1234567}(\vk)&=& \int_p  \left[ l_{31}^2 \vp_{123}^2 +l_{32}^2 \frac{(\vp_{123}\cdot\vp_1)^2}{\vp_1^2}+
  l_{33}^2  \vp_{123}^2 {(\vp_1\cdot\vp_2)^2 \over \vp_1^2 \vp_2^2}+ l_{34}^2   {\vp_{123}\cdot\vp_1\  \vp_{123}\cdot\vp_2 \ \vp_1\cdot\vp_2 \over \vp_1^2 \vp_2^2} \right. \nonumber \\
&+&   l_{35}^2    {(\vp_{123}\cdot\vp_1)^2(\vp_2\cdot\vp_3)^2 \  \over \vp_1^2 \vp_2^2 \vp_3^2}+
 l_{36}^2  \vp_{123}^2  {\vp_{3}\cdot\vp_1\  \vp_1\cdot\vp_2 \  \vp_2\cdot\vp_3 \  \over \vp_1^2 \vp_2^2 \vp_3^2}
\nonumber\\
&+& \left. l_{37}^2   {\vp_{123}\cdot\vp_1\  \vp_1\cdot\vp_2 \  \vp_2\cdot\vp_3\ \vp_{3}\cdot\vp_{123}\  \over \vp_1^2 \vp_2^2 \vp_3^2}\right] \nonumber \\
&\times&   \delta_0(\vp_1)\delta_0(\vp_2) \delta_0(\vp_3).
 \eea
All these terms are local combinations of the first order gravitational potential and thus result in a contribution 
\beq
\langle \delta^{ct3}\delta^{SPT(1)}\rangle |_{1234567} = k^2 {\bar l}_{3}^2 P(k).
\eeq
The contribution in  equation (\ref{ct3}) coming from the third order displacement can also be ignored as the divergence of the third order displacement can be rewritten in terms of local combinations of lower order terms and thus is degenerate with terms we already have. As a result the only non-trivial contributions are those coming from the combination of the second order displacement and the first order one. The resulting contribution to the density is:
\bea
\delta^{ct(3)}(\vk)&=& \int_p  \left[ l_{36}^2 \vp_{123}^2 { {(\vp_3\cdot\vp_{12})^2 \over \vp_3^2 \vp_{12}^2}} (1-{(\vp_1\cdot\vp_2)^2 \over \vp_1^2 \vp_2^2}) + l_{37}^2   {\vp_{123}\cdot\vp_3\  \vp_{123}\cdot\vp_{12} \ \vp_3\cdot\vp_{12} \over \vp_3^2 \vp_{12}^2}(1-{(\vp_1\cdot\vp_2)^2 \over \vp_1^2 \vp_2^2})  \right] \nonumber \\
&\times&   \delta_0(\vp_1)\delta_0(\vp_2) \delta_0(\vp_3).
 \eea
We can define $\vp_3\cdot\vk = p_3 k x$ and $r=p_3/k$ and write
\bea
\langle \delta^{ct(3)}\delta^{SPT(1)}\rangle &=& P(k)  \nonumber \\
&\times& 2 k^3 \int {dr r^2 \over (2 \pi)^2 } P(k r)  \int_{-1}^1 dx  \left[ - k^2 l_{38}^2  {(1-x^2)^2 \over y(r,x)} -  k^2  { l}_{39}^2   {r x (1-x^2)^2 \over y(r,x)}  \right ],
 \eea
where we have removed contributions proportional to $k^2 P(k)$ by using $2rx = 1+r^2 -y(r,x)$ and dropping every term with no inverse power of $y(r,x)$.

\section{IR Resummation}

Motions produced by long wavelength modes are important in smoothing the BAO peak in the correlation function, or suppressing the BAO oscillations in the power spectrum. The effect is governed by a different expansion parameter and is not small. It is possible however to sum these contributions, a technique usually called infra-red resummation \cite{Senatore:2014via,Baldauf:2015xfa}.  This method multiplies the oscillatory part of the power spectrum (the wiggle part $P_\text{w}$) by an exponential damping but leaves the broadband part (the no-wiggle part $P_\text{nw}$) unaffected. The result for the two-loop power spectrum is:
\beq
\begin{split}
P_\text{IR}(k)=&e^{-\Sigma^2_{\epsilon k}k^2}\left[\left(1+\Sigma^2_{\epsilon k}k^2
-\frac{1}{2}\Sigma^4_{\epsilon k}k^4\right)P_{{w}}
+(1+\Sigma^2_{\epsilon k}k^2)P^\text{1loop}_{w}+P^\text{2loop}_{w}\right]\\[10pt]
&+P_{{nw}}+P^\text{1loop}_{nw}+P^\text{2loop}_{nw}\; ,
\end{split}
\eeq
where the linear power spectrum is decomposed as $P=P_w+P_{nw}$, with $P_w$ the BAO wiggle part of the power spectrum; $P^\text{1,2loop}_{nw}$ are the 1-loop or 2-loop power spectra computed using $P_{nw}$ (including their counter terms), and
\beq
P^\text{1,2loop}_{w}= P^\text{1,2loop}-P^\text{1,2loop}_{nw}+\mathcal{O}((P_w/P_{nw})^2),
\eeq
where $P^\text{1,2loop}$ are the full 1- or 2-loop power spectra (again including their counter-terms). In the following we neglect the quadratic corrections proportional to $P_w$. We have also introduced
\beq
\Sigma^2_\Lambda=\frac{1}{3}\int_0^\Lambda \frac{d^3 q}{(2\pi)^3} \frac{P(q)}{q^2}\bigl[1-j_1(q r_\text{BAO})+2j_2(q r_\text{BAO})\bigr] \;,
\eeq
with $j_n$ being the $n^\text{th}$ order spherical Bessel function. We only consider the smoothing due to motions arising from scales much larger than the scale under consideration, for definiteness we choose $\epsilon=1/2$.

\section{Two-loop power spectrum counter terms in $\Lambda$CDM}

\begin{figure}
\centering
\includegraphics[width=0.5\textwidth]{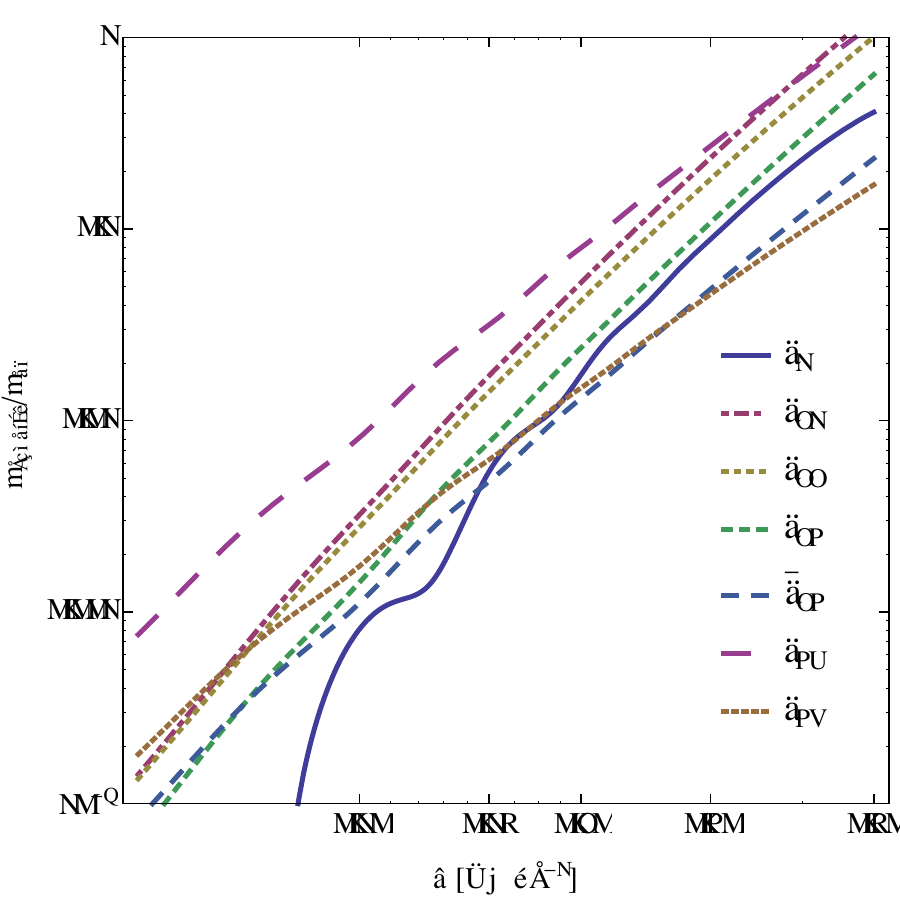}
\includegraphics[width=0.47\textwidth]{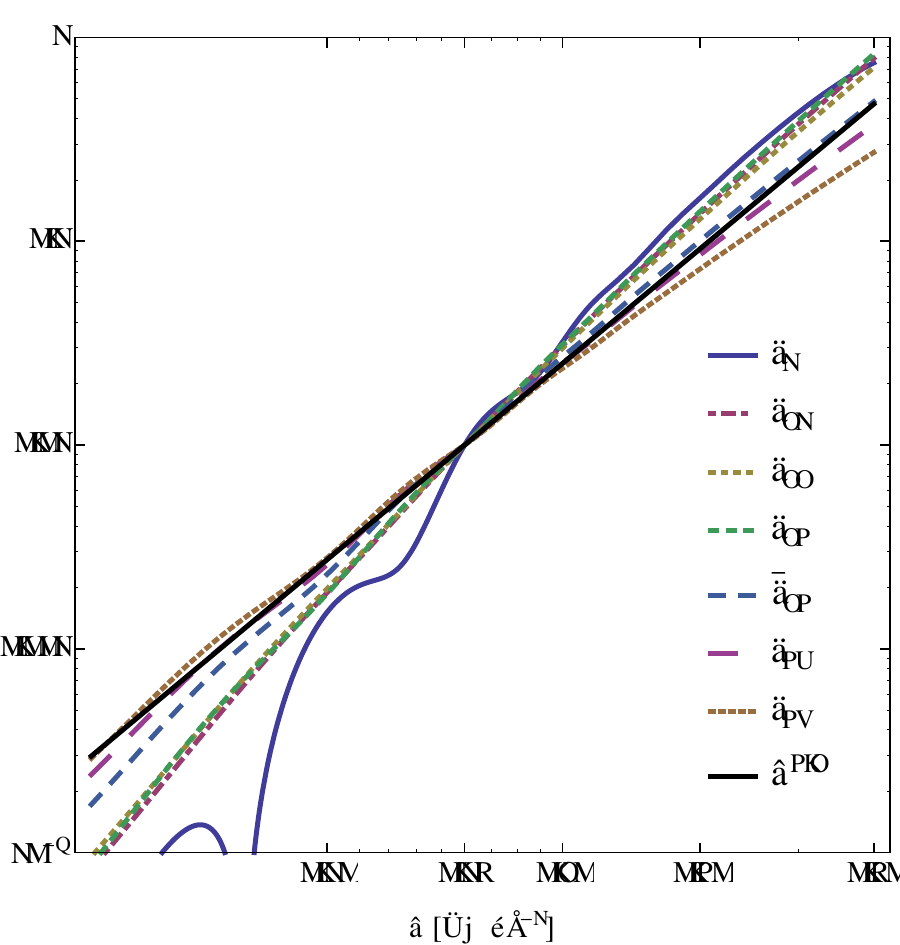}
\caption{Left: Two loop counter term contributions for $l^2=1\ (h^{-1} \rm{Mpc})^2$. Right: Two loop counter term contributions normalized so that they are 1\% of the linear power spectrum at $k=0.15 \ h \rm{Mpc}^{-1}$. We also show a representative power law that roughly captures the scaling of these terms.}
\label{fig:plotleq1}
\end{figure}

Figure \ref{fig:plotleq1} shows the IR resumed two loop counter terms normalized so that all the length scales are the same $l^2_{n}=1 \ h^{-2}\rm{Mpc}^2$. This normalization is the one found for the linear counter term when fitting a one loop calculation of the power spectrum to the measurements of dark matter simulation ({\it eg.} \cite{Baldauf:2015aha}). More detailed comparison with numerical simulations including higher order terms seem to indicate that the second order counter terms computed here contribute at the percent level at $k=0.15 \ h\rm{Mpc}^{-1}$ \cite{Foreman:2015lca,Baldauf:2015aha}. These two estimates are consistent. 

Figure \ref{fig:plotleq1} shows that the shape of the various terms is similar, but not identical. This can be seen more clearly in the bottom panel where we normalized all the terms to make a 1\% contribution around $k=0.15 \ h\rm{Mpc}^{-1}$. The fact that all the shapes are similar implies that if one is interested only in fits of the dark matter power spectrum, the different terms are going to be almost degenerate. Equivalently one expects to be able to describe the effect of all these terms by keeping only one or two. This was indeed the conclusion of  \cite{Foreman:2015lca}.

One should point out that the only reason why the various counter terms are not identical is that the power spectrum of matter fluctuation in $\Lambda$CDM is not a power law. If it were, all the terms we have computed would have the same scale dependence, $k^{2 n + 5}$ where $n$ is the slope of the density power spectrum. Around $k\sim 0.2  h\rm{Mpc}^{-1}$ the slope of the no-wiggle power spectrum is around $n\sim-1.8$. The bottom panel of figure \ref{fig:plotleq1} shows the predicted power law for these slope, it captures the typical scaling of the new counter terms reasonably well. These representative power law is shallower than that of the contribution from the next higher derivative operator that scales as $k^4 P\propto k^{n+4}$. However this term is not so much steeper than the steepest of the counter terms we have computed. For example properly normalized this higher derivative contribution and the term proportional to $l_{21}^2$ differ by less that 30\% in the range $0.05  h\rm{Mpc}^{-1} < k < 0.4 h\rm{Mpc}^{-1}$.

\begin{figure}[t!]
\centering
\includegraphics[width=0.4\textwidth]{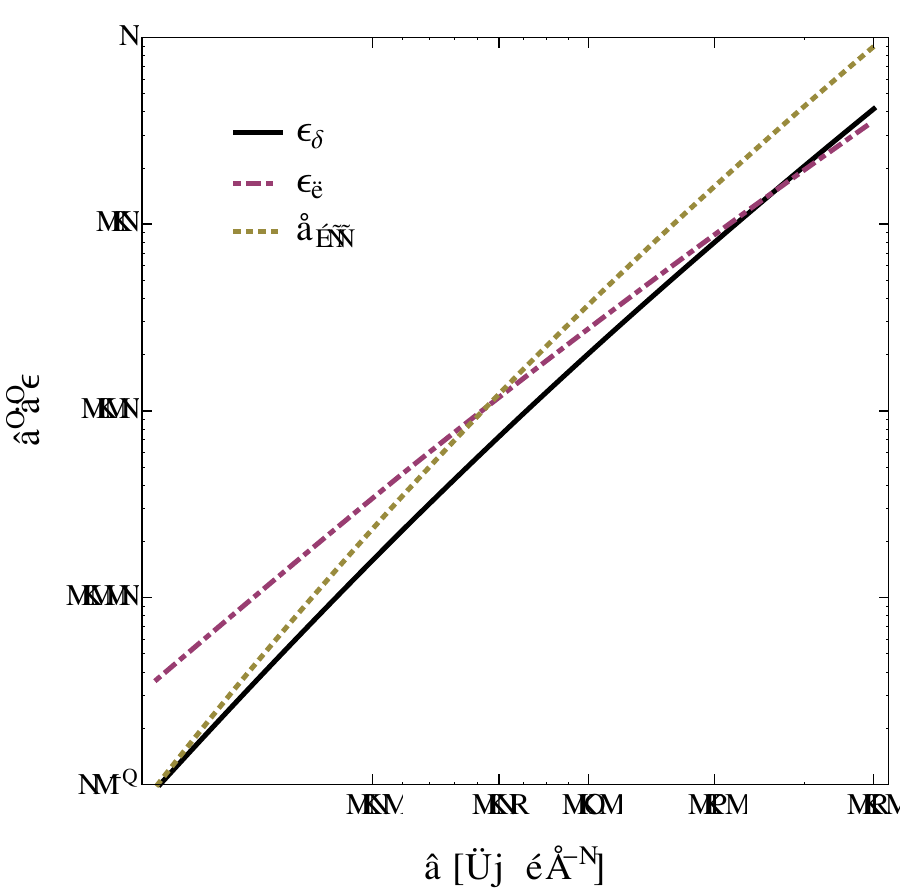}
\caption{The size of the two loop counter terms are controlled by  $k^2 l^2 \epsilon_\delta$,  $n_{eff} k^2 l^2 \epsilon_\delta$.  For the plot we have chosen $l^2=1 \ h^{-2}\rm{Mpc}^2$ and $n_{eff}$ is computed for the power spectrum without BAO wiggles. }
\label{fig:plotepsilons}
\end{figure}

\begin{figure}
\centering
\includegraphics[width=0.56\textwidth]{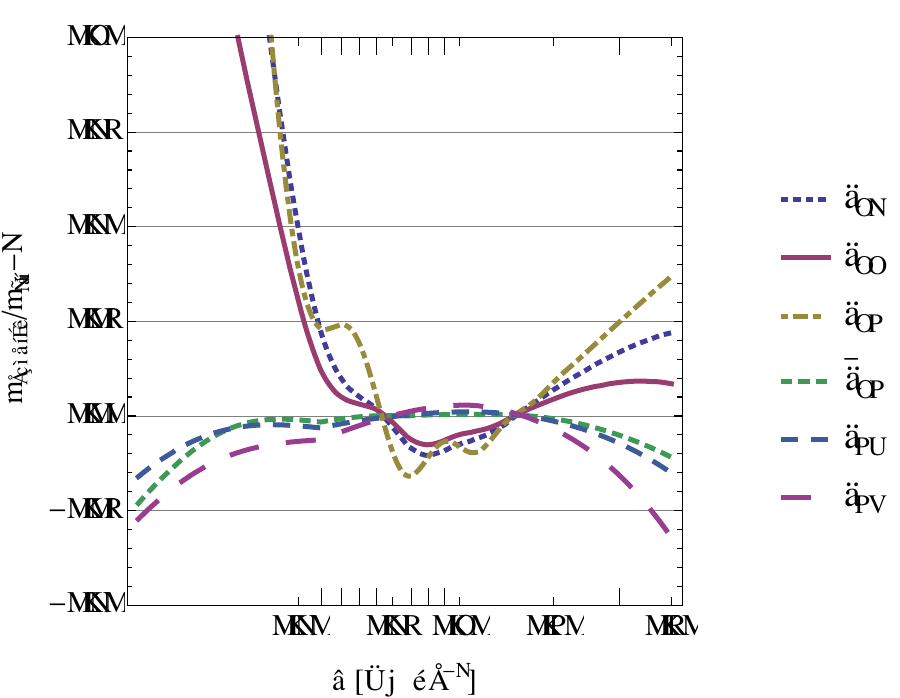}
\includegraphics[width=0.4\textwidth]{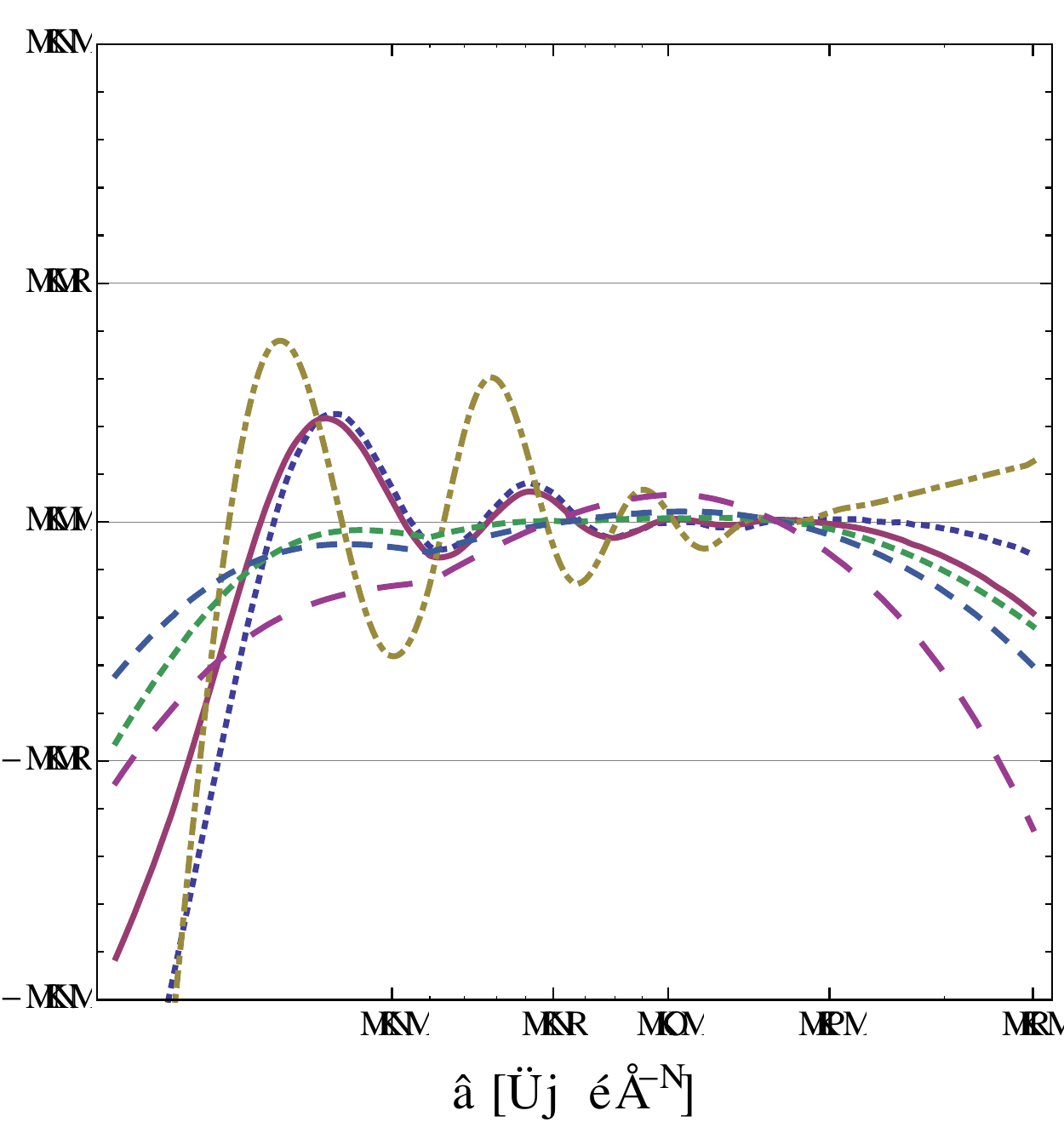}
\caption{Left:  Fit to the five two loop counter terms using a linear combination of $k^2 P \epsilon_\delta$ and  $k^2 P \epsilon_s$ . Right: adding a term proportional to  $k^2 n_{eff} P \epsilon_\delta$.}
\label{fig:fit}
\end{figure}

One can get some further intuition as to what happens in $\Lambda$CDM by looking at the contribution to the momentum integrals coming for the various counter terms from modes $q\ll k$ and $q\gg k$. These contributions are controlled by different parameters. The amplitude of the contribution from modes with $q\gg k$ is determined $k^2 l^2 \epsilon_s$ while the effect of modes $q\ll k$ depends on two separate parameters, $k^2 l^2 \epsilon_\delta$ and $n_{eff} k^2 l^2 \epsilon_\delta$ defined as 
\bea
\epsilon_s &=& k^2 \int_{q \gg k} {d^3 q \over (2\pi)^3} {P(q) \over q^2}\nonumber \\
 \epsilon_\delta &=& \int_{q \ll k} {d^3 q \over (2\pi)^3} P(q)  \nonumber \\ 
n_{eff} &=& {d \log P \over d\log k}.
\eea

The three parameter controlling the size of the IR and UV contributions to the integrals  are shown in figure \ref{fig:plotepsilons}. Their shape is not identical. The contribution from each of these pieces to the various counter terms we computed in this paper is different. This accounts for the slight difference in the shapes of the terms we calculated. For example the two IR parameters ( $k^2 l^2 \epsilon_\delta$ and $n_{eff} k^2 l^2 \epsilon_\delta$) give the biggest contribution to the quadratic counter terms over the scales of interest while the $\epsilon_s$ is the biggest contribution to the cubic ones. Another difference between the terms is that the $2-2$ type integrals (when a quadratic term is correlated with the quadratic piece of the SPT result) does not show residual wiggles. The $1-3$ type integrals (when a cubic term is correlated with the linear piece of the SPT result) of course does as it is proportional to the linear power spectrum (before IR resummation). We note that the counter term proportional to $l_{23}^2$ contains a $1-3$ piece, which is equal to the term proportional to ${\bar l}_{23}^2$ which as defined does not have a 2-2 contribution. So both the $l_{23}^2$ and ${\bar l}_{23}^2$ have some wiggles.

The higher order counter term that derives from the linear counter terms is also parametrically different. This is due to two facts, its IR contribution has an additional dependence on  $\alpha_{eff} k^2 l^2 \epsilon_\delta$ where the running of the spectral index is $\alpha_{eff}= {d n_{eff}  \over d\log k}$. Furthermore the UV and IR pieces are more comparable leading to a partial cancellations that changes the final shape. 

It is instructive to see how well the counter terms can be fit over the range of interest as a combination of their limits $k^2 P \epsilon_\delta$, $k^2  P \epsilon_s$ and $k^2 n_{eff} P \epsilon_\delta$, where for the slope we have taken the slope of the power spectrum without the BAO wiggles. This is shown in figure \ref{fig:fit}. 
Even the combination of the first two terms does a very good job in approximating the different shapes, getting residuals at a few percent level. Given that at $k=0.2 \ h\rm{Mpc}^{-1}$ the counter terms we have studied  contribute at the few percent level, percent level residuals should be sufficient to compare with existing simulations.  Most of the difference is seen for the terms dominated by the IR contribution. Our approximation in terms of $n_{eff}$ is probably too crude. In any case, due to their steep slope, the range of scales over which these counter terms make a significant contribution but still the system can be understood perturbatively is quite limited so only for very precise simulations and or observations would the difference in shape be important. At present one or two additional parameters should suffice.

\acknowledgements
We thank Tobias Baldauf and Marko Simonovi\'c for useful discussions. M.Z. is supported in part by the NSF grants  PHY-1213563, AST-1409709 and PHY-1521097. M.M. acknowledges support from NSF Grants PHY-1314311 and PHY-1521097.

\bibliographystyle{utphys}
\bibliography{spt-lag}

\end{document}